\documentclass{article}

\usepackage{paperstyle}
\usepackage[utf8]{inputenc}
\usepackage[T1]{fontenc}
\usepackage{graphicx}
\usepackage{hyperref}
\usepackage{url}
\usepackage{booktabs}
\usepackage{amsfonts}
\usepackage{nicefrac}
\usepackage{microtype}
\usepackage{lipsum}
\usepackage{fancyhdr}
\usepackage{graphicx}
\usepackage{xcolor}
\usepackage{makecell}
\usepackage{multirow}

\title{Bypassing Document Ingestion:\\ An MCP Approach to Financial Q\&A}

\author{
  Sasan Mansouri \\
  University of Groningen \\
  Groningen, Netherlands \\
  \texttt{s.mansouri@rug.nl} \\
  \And
  Edoardo Pilla, Mark Wahrenburg \\
  Goethe University Frankfurt \\
  Frankfurt am Main, Germany \\
  \texttt{\{pilla, wahrenburg\}@finance.uni-frankfurt.de} \\
  \And
  Fabian Woebbeking \\
  Halle Institute for Economic Research (IWH) \\
  and Martin Luther University Halle-Wittenberg \\
  Halle (Saale), Germany \\
  \texttt{fabian.woebbeking@iwh-halle.de} \\
}

\begin{document}
\maketitle

\begin{abstract}

Answering financial questions is often treated as an information retrieval problem. In practice, however, much of the relevant information is already available in curated vendor systems, especially for quantitative analysis. We study whether, and under which conditions, Model Context Protocol (MCP) offers a more reliable alternative to standard retrieval-augmented generation (RAG) by allowing large language models (LLMs) to interact directly with data rather than relying on document ingestion and chunk retrieval. We test this by building a custom MCP server that exposes LSEG APIs as tools and evaluating it on the FinDER benchmark. The approach performs particularly well on the Financials subset, achieving up to 80.4\% accuracy on multi-step numerical questions when relevant context is retrieved. The paper thus provides both a baseline for MCP-based financial question answering (QA) and evidence on where this approach breaks down, such as for questions requiring qualitative or document-specific context. Overall, direct access to curated data is a lightweight and effective alternative to document-centric RAG for quantitative financial QA, but not a substitute for all financial QA tasks.

\end{abstract}

\keywords{Financial Benchmarks \and Information Extraction \and Large Language Models \and LSEG \and Model Context Protocol}

\setcounter{footnote}{0}

\section{Introduction}

In finance, information is not scarce. The challenge is to use it reliably. Recent advances in LLMs have made textual information easier to process, but they have also encouraged the construction of ad hoc retrieval pipelines around information that is often already available in curated form. In a setting where accuracy is critical, this is not only inefficient but also potentially risky, as LLM- and RAG-based systems remain vulnerable to retrieval and generation errors, and non-transparent failures. We examine whether MCP-based systems, which enable direct interaction with curated financial data, provide a more reliable approach on finance-specific benchmarks. In fact, when it comes to artificial intelligence (AI) adoption, the financial domain has shown promising potential in different subfields, including accounting and reporting. However, acceptance of the technology largely depends on the degree of accuracy, security and traceability that the models can achieve. The tendency of LLMs to provide factually inaccurate statements, an issue commonly referred to as \textit{hallucination} in the literature, represents one of the first major obstacles to generalized AI adoption in the financial sector, where maximizing factual accuracy is paramount. At the same time, the knowledge that models can inject in answer generation through the so-called \textit{look-ahead bias} limits the extent to which practitioners decide to implement generative AI in their respective workflows, especially those where generation should be exclusively based on a predetermined information set.\cite{kandpal2023longtail,mallen2023parametrics,levy2024bias,sarkar2024bias}

RAG has been one of the solutions that were initially researched and proposed to tackle the problem of hallucination: by providing the relevant context to LLMs, an answer can be generated that is grounded in truth and references, as defined within the respective knowledge base. This technique therefore serves to reduce the degree of factually inaccurate and incorrect answers generated by LLMs, and entails extracting the most relevant pieces of information from a document to answer the question at hand.\footnote{Implementing guardrails in the generation pipeline to prevent LLMs from hallucinating is in itself a solution to the issue, but most often not the optimal one, given that the user is ultimately interested in receiving the correct answer to its query, rather than observing a non-answer or an apology for the inability of the LLM to generate a factually accurate output.} However, RAG pipelines hinge on careful and well-designed document parsing, chunking of the resulting knowledge base, and subsequent indexing of chunks by means of embedding vectors, that capture and maintain semantic relationships across the knowledge base and are usually hosted in vector stores, steps that often prove to be challenging if the document layout is not elementary. Furthermore, parts of the infrastructure, such as optical character recognition (OCR), which is normally used to enable usage of knowledge bases that are not originally available in a machine-readable format, can quickly become serious bottlenecks when trying to ensure scalability of the system.\cite{lewis2020rag}

Tables and charts, extremely important collections of quantitative data points for practitioners in the financial industry, represent an added layer of complexity for standard RAG pipelines, as their layout is often not as straightforward as that of textual paragraphs, and thus often lead to reduced performance of the retrieval system due to erroneous or incomplete parsing. Robust retrieval pipelines are especially necessary when attempting to embed LLMs in specific subfields of the financial industry, such as fraud detection, due to the nature of the data involved. However, building specialized infrastructures capable of correctly ingesting and processing tabular data is associated with overhead costs that justify seeking alternative solutions.\cite{tan2025table}

Towards the end of 2024, MCP has been released as a unified framework aimed at standardizing communication between LLMs and external services. One of the possible use cases for MCP, building on tool and function calling features, which had been implemented by main model providers prior to MCP coming into existence, is efficiently and easily connecting LLMs and external data vendors, in order to transfer data that the provider supplies to the model as context, for the purpose of generating an answer grounded in evidence, which was previously collected and disseminated by reliable sources. This represents an alternative approach to standard RAG, and entirely removes the necessity to set up an infrastructure that ingests and processes documents. Instead, context is dynamically selected and passed to the generator model, effectively delegating to LLMs the management of their context window. In this regard, MCP is becoming a widespread standard in ensuring seamless, transparent communication between LLMs and data providers, thus systematically increasing trust that practitioners place in AI within the financial industry, as it can now be powered with knowledge supplied by reliable vendors. To this end, it is therefore important to test and measure performance of LLMs when the context that they receive stems from such sources, rather than from financial reports and statements.\cite{anthropic2024mcp,challapally2025nanda}

The paper aims at proposing precisely that, and it contributes to the literature by releasing a custom MCP server that is ready to use with an appropriate license\footnote{In October 2025, LSEG has announced the release of a proprietary MCP server that is currently available to end users. However, coverage ensured by the defined tools is, as of March 2026, focused on the currency and fixed income markets, whereas the custom MCP server developed here mainly targets corporate and reference data.\cite{hussain2025lseg}}, along with a study on performance of external data provider-powered LLMs on financial QA datasets, an area of research that is still underexplored. This effort also highlights the importance and lack of human-curated datasets acting as testbeds for further MCP-based solutions spanning the financial domain, given that most of the available benchmarks contain questions that are specifically designed and tailored to prompt an answer that is generated after accessing a document, rather than through tool execution.

The main results emphasize the effectiveness of an MCP approach in handling quantitative questions pertaining to financial analysis, with an unconditional overall accuracy of 69.7\% obtained in the \textit{Financials} subset of the reference dataset, and a maximum accuracy of 80.4\% obtained in the group of questions for which the system is able to extract relevant context, and involving multiple quantitative steps of potentially differing nature, determined within the aforementioned sample. Robustness of the pipeline in handling quantitative questions is furthermore confirmed by the systematically higher \textit{Context Relevance} and \textit{Response Groundedness} scores displayed within the subset of financial analysis-related answers, with average values respectively amounting to 72.5\% and 90.4\%, as opposed to the remainder of the reference dataset, whose averages equal 20.5\% and 69.7\%. These results are achieved without the complexity derived from parsing tables and charts, which most often contain the quantitative data necessary to accurately answer when attempting to extract information directly from financial reports and commonly represent a challenge for standard RAG architectures.

In fact, for standard RAG architectures, accessing tables and charts is fundamental in order to extract the quantitative figures that constitute the context to be transferred over to LLMs so that a factually correct answer can be generated, as opposed to textual paragraphs within the respective document, which mostly focus on qualitative information. An MCP approach thus configures a lightweight alternative to this and proves to be a particularly effective workaround to standard RAG in the financial domain, since validated and clean quantitative data are already available and efficiently disseminated by established providers. Given the degree of importance that quantitative figures display in analysis and decision-making within the financial industry, such option represents an effective way to handle quantitative questions that can be easily embedded in agentic pipelines to be orchestrated and selected dynamically when receiving questions of suitable nature by the same end user, so that LLMs powered with access to external data providers are chosen for incoming quantitative queries, whereas standard RAG enters into action in responding to inquiries concerning qualitative information.

The remainder of the paper is organized as follows: Section 2 provides an outline of the related literature. Section 3 introduces the methodology behind the study and provides further insight into the system architecture. Section 4 displays the main results stemming from benchmarking runs and the respective robustness checks. Finally, Section 5 draws the main conclusions, discloses the existing limitations, and sets forth a set of initiatives for further research and development.

\section{Related work}

\subsection{Corporate financial analysis}

Academic efforts in the area have spanned several decades and dimensions, and have shifted from less structured attempts that were prone to individual idiosyncrasies, to more rigorous and grounded in statistical evidence as data have become available at scale to researchers and practitioners in the industry. Particular attention has been attributed to the field of corporate distress and bankruptcy, and more broadly to the concept of risk, due to the severe financial impact that adverse events can determine for different stakeholders.\cite{altman1968bankrupt,altman1984business,almamy2016zscore}

Quantitative data points routinely play a fundamental role in enabling extensive and systematic assessment of corporate health, summarizing and synthesizing the nature of its condition within the economic system, and the precision with which numbers are reported can enhance or hinder any analysis that builds upon them, deeply influencing the efficiency of information transmission in financial markets when attempting to unveil dynamics that are implied in the respective accounting choices. Conversely, the advances in natural language processing (NLP) have enabled to paint a comprehensive picture of reporting corporates by taking into account qualitative information and forward-looking statements that company officials routinely include in periodic disclosures.\cite{beaver2012accounting, deangelo1994accounting,bozanic2018forward,barth2022nonanswers}

Additionally, clear and accurate reporting of data commonly contained in financial statements and disclosures is instrumental in allowing extensive analysis of firm behavior over time, a crucial aspect in detecting and quantifying the extent of earnings smoothing practices, potentially originating from proximity to distress and violation of financial obligations. Research has also shown that reliably accessing compensation data, normally included in structured corporate disclosures, helps in transparently assessing any potential conflict of interest that may arise and result in different accounting decisions that managers make to increase their compensation.\cite{dechow2012earnings,franz2014earnings,graham2005reporting, healy1985bonus}

\subsection{Financial QA datasets}

Research and measurement of AI capabilities in the financial domain have commenced as soon as the technology became widely available. Early work on financial QA benchmarks predates the release of modern LLMs to the general public and was originally carried out to gauge performance of more traditional NLP algorithms, but datasets such as FinQA and TAT-QA have subsequently become standard for assessing the behavior of LLMs on complex, multi-step numerical reasoning tasks that commonly constitute typical workflows in the financial industry.\cite{chen2021finqa,zhu2021tatqa}

More recently, FinanceBench and FinDER have also been published as human-annotated sources of question-answer pairs whose evidence is grounded in the respective knowledge bases, covering major companies that constitute highly liquid market indexes such as the American S\&P500. These benchmarks are characterized by a correspondence between answer reference, or ground truth, and knowledge base index, such as the page number at which the relevant piece of information required to answer a given question can be found inside the respective document, which is typically a 10-K or 10-Q report. Moreover, these datasets enable researchers to condition on task features such as the type of reasoning that a task entails, or the specific subject area to which a question refers. This is particularly useful as it allows to achieve a deeper understanding of relationships between model performance and task type, while quickly highlighting critical areas where LLMs do not perform in a satisfactory manner.\cite{islam2023finbench,choi2025finder}

\subsection{Retrieval-augmented generation}

Recent developments in the field of textual analysis and AI have crystallized in 2020 with the release of GPT-3, a model suited for natural language generation at scale, deriving from work in the area of NLP that introduced the Transformer architecture, whose root mechanism is attention. Subsequently, services based on access to LLMs through user interfaces, such as ChatGPT in 2022, enabled the general public to incorporate LLMs into their daily workflows in several industries.\cite{bahdanau2014attention,brown2020gpt3,kim2017attention,vaswani2017att}

RAG has subsequently emerged as a method to reduce or eliminate the degree of hallucination that LLMs display across tasks in different domains, namely the tendency of LLMs to generate an output that can seem plausible, but is not grounded in factual evidence and truth. Hallucinations are, among other reasons, caused by training cutoff dates of the models, that effectively truncate the knowledge available in-house to the generator, and the implied inability of LLMs to keep up with punctual, real-time updates, which was particularly problematic prior to the advent of tools and function calling, that have subsequently equipped LLMs with capabilities such as internet browsing prior to final answer generation. However, in specific domains such as the financial one, a relevant culprit when observing hallucinations lies in the absence or reduced presence of data that should be picked up and used by the model to answer a domain-specific question within the corpora that are fed to LLMs during their respective pre-training phases, which leads these data to being labeled as \textit{long-tail knowledge} in commonly referenced taxonomies.\cite{kandpal2023longtail,sarmah2023rag,gao2024rag,huang2024hal}

RAG can therefore serve as an alternative to relying on knowledge that is embedded into LLMs during pre-training and training in order to generate a factually accurate and correct answer and hinges, in its standard implementation that extracts knowledge from unstructured documents, on efficient ingestion of the relevant knowledge base, in addition to accurate chunking, that is mostly defined at the document paragraph-level, and often refined through specific techniques such as chunk re-ranking, metadata annotations and vector embedding fine-tuning algorithms, or more recently by leveraging the improved self-reflective skills that the latest generation of models exhibits.\cite{setty2024rag,yepes2024chunk,ferrazzi2026rag}

With the goal in mind of streamlining RAG pipelines and minimizing human intervention and bias in the evaluation process, the \textit{LLM-as-a-Judge} paradigm has emerged. Recent research in the field focused on overcoming systematic errors and inconsistencies through joint consideration of scores generated by multiple models, either as a statically or dynamically computed weighted average based on each evaluator model's reliability, or through more complex feedback loops determined by evaluator agentic systems. The \textit{Retrieval Augmented Generation Assessment} framework (RAGAS) has been proposed and is gradually being adopted as a standard in the industry and in research environments for automated RAG system evaluation based on an LLM-as-a-Judge approach. This framework offers a suite of evaluation metrics that pertain to the retrieval and generation components, allowing a comprehensive evaluation of the developed RAG pipeline along all its dimensions.\cite{gu2025llmasjudge,zhuge2024llmasjudge,corradaemmanuel2025llmasjudge,li2025llmasjudge,es2025ragas}

\subsection{Model Context Protocol}

Academic efforts on MCP have begun shortly after its release to the general public in November 2024. Initial studies spanned several domains across the literature, and mostly focused on addressing its impact over cybersecurity and corporate risk, by proposing workarounds and solutions that are feasible in the enterprise environment.\cite{errico2025mcp,hou2025mcp,wang2025mcpguard}

In parallel, a few benchmarks have been developed to assess performance of agentic pipelines operating within MCP on a diverse set of tasks, ranging from straightforward to sensibly complex, in order to provide the community with standardized frameworks serving the purpose of reliably constructing AI agent-powered systems that can leverage knowledge and data stemming from different providers and vendors.\cite{guo2025mcp,wang2025mcpbench}

Capabilities of MCP within the financial field have, however, only received limited attention and ample scope for exploration exists, given the sheer quantity of data efficiently provided and disseminated by commercial and non-commercial providers, along with the vast range of applications in the industry that could benefit from further streamlining and automation made possible by AI. In fact, while other studies have focused on equipping LLMs with knowledge stemming from external data vendors, benchmarking their performance on established financial QA datasets still represents an underexplored area.\cite{bhandari2025mcp,zeng2025mcp}

\section{Methodology}

\subsection{System architecture}

The engine that powers LLMs with knowledge served by LSEG is, at its core, an MCP server that exposes API endpoints, accessible to any user who has a valid LSEG Workspace license, as MCP tools that the models can choose to employ in order to generate an answer to any given question.\footnote{The infrastructure is publicly available at https://github.com/edoardopilla/lseg\_llm\_mcp} Figure 1 offers a high-level overview of the architecture and captures the interactions that occur across the underlying modules, whereas a comprehensive list of tools, along with a summary of their designed behavior, is contained in Table 1.\cite{anthropic2024mcp}

\begin{figure}[h!]
    \centering
    \includegraphics[width=.9\columnwidth]{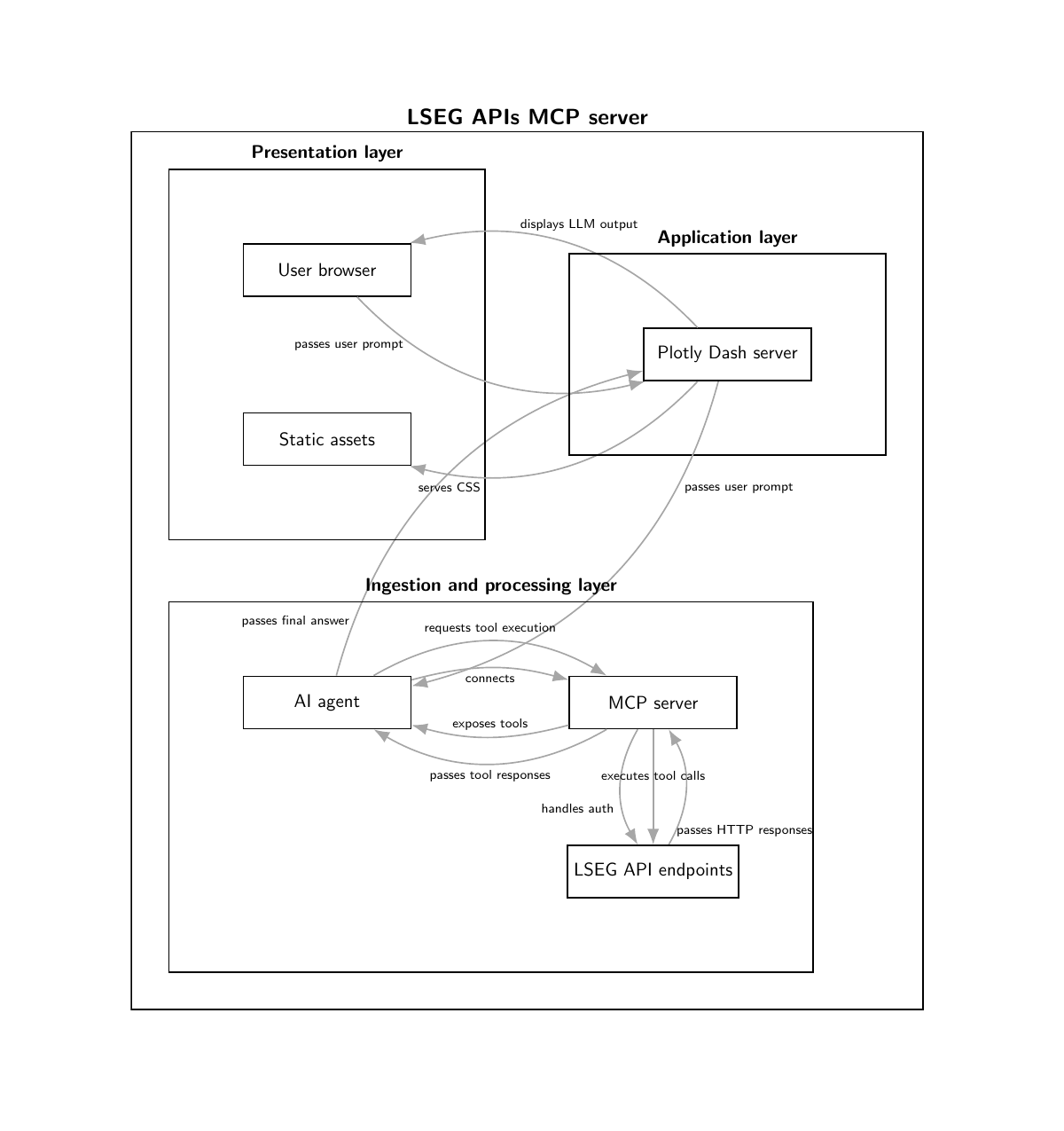}
    \caption{\textit{LSEG APIs MCP server system architecture diagram.}}
\end{figure}

An extensive set of metadata is automatically generated for each tool execution and is readily accessible, easing the analysis and enabling a fully transparent inspection of generated answers for the purpose of excluding hallucinations and ensuring that the final model response is directly linked to the ground truth, as it is provided by LSEG. A simple implementation of structured output eventually ensures that final answers are extracted from the potentially longer full output that the generator model creates, an additional layer that allows to swiftly determine alignment between ground truth and generated response when evaluating system behavior and that is especially useful when observing quantitative figures.

\begin{table}[h!]
\small
\begin{center}
\caption{Implemented MCP tools.}
\begin{tabular}{ccc}
\hline
\textbf{Name} & \textbf{Arguments} & \textbf{Description} \\
\hline
\textit{get\_acquisitions} & \makecell{\textit{comp\_name}\\\textit{end\_date}\\\textit{start\_date}} & Finds relevant M\&A data \\
\hline
\textit{get\_balancesheet\_statement} & \makecell{\textit{comp\_name}\\\textit{period}\\\textit{scale}} & Retrieves relevant balance sheet statement \\
\hline
\textit{get\_business\_segments} & \makecell{\textit{comp\_name}\\\textit{period}\\\textit{scale}} & Retrieves relevant business segment revenue data \\
\hline
\textit{get\_capital\_structure} & \makecell{\textit{comp\_name}\\\textit{period}\\\textit{scale}} & Retrieves relevant capital structure data \\
\hline
\textit{get\_cashflow\_statement} & \makecell{\textit{comp\_name}\\\textit{period}\\\textit{scale}} & Retrieves relevant cash flow statement \\
\hline
\textit{get\_earningscall\_transcript} & \makecell{\textit{comp\_name}\\\textit{period}} & Outputs relevant earnings call transcript \\
\hline
\textit{get\_geographic\_segments} & \makecell{\textit{comp\_name}\\\textit{period}\\\textit{scale}} & Retrieves relevant geographic segment revenue data \\
\hline
\textit{get\_income\_statement} & \makecell{\textit{comp\_name}\\\textit{period}\\\textit{scale}} & Retrieves relevant income statement \\
\hline
\textit{get\_operating\_metrics} & \makecell{\textit{comp\_name}\\\textit{period}\\\textit{scale}} & Retrieves relevant operating metrics, including physical ones \\
\hline
\textit{get\_pension\_plan} & \makecell{\textit{comp\_name}\\\textit{period}\\\textit{scale}} & Retrieves relevant pension plan data \\
\hline
\textit{get\_product\_segments} & \makecell{\textit{comp\_name}\\\textit{period}\\\textit{scale}} & Retrieves relevant product segment revenue data \\
\hline

\end{tabular}
\end{center}
\end{table}

\subsection{Reference dataset}

FinDER represents the main testbed used for assessing performance of LSEG-powered LLMs due to its size, which enables to draw meaningful conclusions, and due to the shape of its questions, that are more adherent to a practitioner's behavior compared to other benchmarks available for research. A separate preprocessing step slightly tweaks the original dataset and aligns nomenclature for convenience. Impact on the initial distribution of questions by type is virtually absent and only generated by step 3, therefore the conclusions drawn within the study can be safely applied to the original dataset. Details concerning preprocessing are collected below for reference:\cite{choi2025finder}

\begin{enumerate}
    \item Replace author-defined value \textit{Subtract} with \textit{Subtraction} in \textit{type} variable. This ensures consistency with the labels provided by the authors, since the \textit{Subtract} label never appears in the paper, therefore reconciling the number of tasks classified as belonging to the \textit{Subtraction} type in the dataset with what is reported in the paper, namely 119.
    \item Fill missing entries with value \textit{No ground truth provided by the authors.} in \textit{answer} variable.
    \item Include rows \textit{8dc5ccdd} and \textit{2dba4bde} in the \textit{reasoning} category by switching the respective values from \textit{False} to \textit{True}. This ensures consistency between the author-defined \textit{type} and \textit{reasoning} variables, and reconciles the number of tasks classified as belonging to the \textit{reasoning} category with what is reported in the paper, namely 883.
    \item Map author-defined \textit{type} variable to unique reasoning categories, namely \textit{Information extraction}, \textit{Logical reasoning} and \textit{Numerical reasoning}, with the latter divided further into the original values \textit{Addition}, \textit{Compositional}, \textit{Division}, \textit{Multiplication} and \textit{Subtraction} as determined by the authors. This allows for a clear correspondence between tasks of different nature, which is otherwise only implicitly defined in the dataset.
    \item Rename \textit{type} and \textit{text} variables to \textit{question\_reasoning} and \textit{question}, respectively. This is performed for cosmetic purposes and does not influence the interpretation of results discussed in the following Sections in any way.
\end{enumerate}

Keeping in mind the goal of testing the behavior of the system on questions involving calculations based on data normally available in tabular format within the relevant knowledge base, a simple exploratory analysis enables to determine that the focus should be placed on one subset of the reference dataset in particular: the \textit{Financials} questions. Proxying for the nature of tasks contained within by means of average numeric density quickly allows to notice that, among 5703 questions contained in the full dataset, the 990 pertaining to financial analysis represent the most suitable pool over which the infrastructure can be appropriately tested, as shown in Figure 2. The results of the exploratory analysis also help to ascertain that the \textit{Accounting} subset does not, despite what the label would suggest at first glance, contain a sizable amount of numbers, as the main sources of references are textual paragraphs within the respective financial reports, rather than tabular data concerning financial statements or other purely quantitative measures.\cite{choi2025finder}\footnote{The pattern shown here is robust to common string cleaning procedures, including but not limited to deleting dataset-specific carriage returns identified by \textit{\_X000D\_} or replacing years with non-numeric placeholders, which do inflate the number count, but otherwise do not change the message conveyed in Figure 2.}

\begin{figure}[h!]
    \centering
    \includegraphics[width=.80\columnwidth]{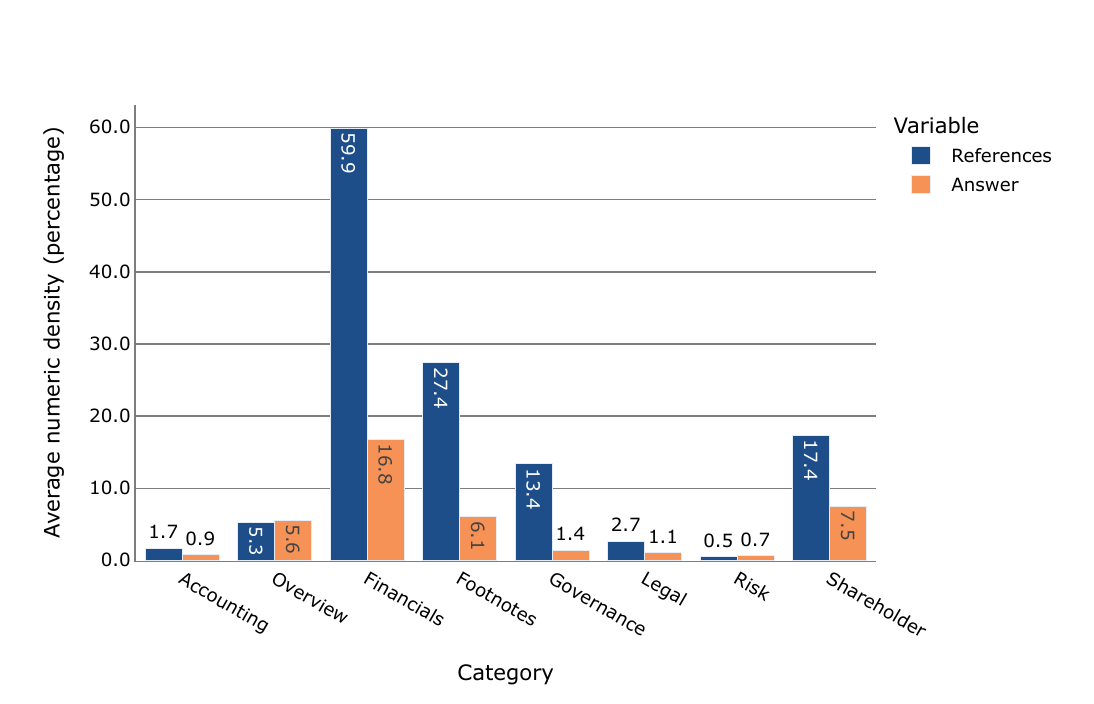}
    \caption{\textit{FinDER average numeric density across answer and references by category. Carriage returns are removed from the Answer variable to avoid inflating the underlying number counts. Numeric density is defined at the granular level as the ratio between number and non-space substring occurrences.}}
\end{figure}

The fact that references within the \textit{Financials} subset contain the highest number occurrences on average hints at another implication that renders this section of the dataset the most suitable for system testing and evaluation, namely that they are rooted in financial statements and other data typically appearing in tabular format within the relevant knowledge base, rather than in qualitative data. Thus, external providers can be most effective in delivering relevant data points to the generator model through MCP-wrapped tools. Focus is therefore shifted to discriminating between those questions for which the system sources data that are most often contained in tables within the respective financial report and would therefore, in the absence of external data providers, only be available conditional on effective ingestion and chunking in a standard RAG pipeline, and those for which data needed to answer mostly appear within the main body of the related financial report, thus not requiring to rely on particularly convoluted ingestion and chunking pipelines. An interesting observation that can be made is that data that are not organized in a tabular fashion, such as qualitative information and forward-looking statements, are not necessarily stored in a structured way among data providers, implying that sustaining the overhead cost of developing an ingestion and chunking pipeline is justified in these cases.\cite{choi2025finder}\footnote{LSEG does offer systematic access to unstructured content stemming from 10-K and 10-Q reports through the proprietary Filings API which, however, is not included in the standard Workspace license.}

Another dimension that can be explored to confirm that the \textit{Financials} subset is the ideal testing ground for the system developed in the study is the average number of forward-looking words that appear within each category of the dataset. Figure 3 shows that questions pertaining to financial analysis exhibit systematically less terms that normally characterize statements advanced by the respective company’s directors and executives, or explanations and insights that build on financial statement figures but are potentially subject to multiple interpretations. Additionally, the references also score lowest here, further proving that the context stored within is closest to tabular data, rather than textual paragraphs.\cite{bozanic2018forward,barth2022nonanswers,choi2025finder}\footnote{The same pattern is visible among the original questions, but not displayed here for cosmetic purposes as the respective averages are, except for the \textit{Risk} category, always lower than 1.}

\begin{figure}[h!]
    \centering
    \includegraphics[width=.80\columnwidth]{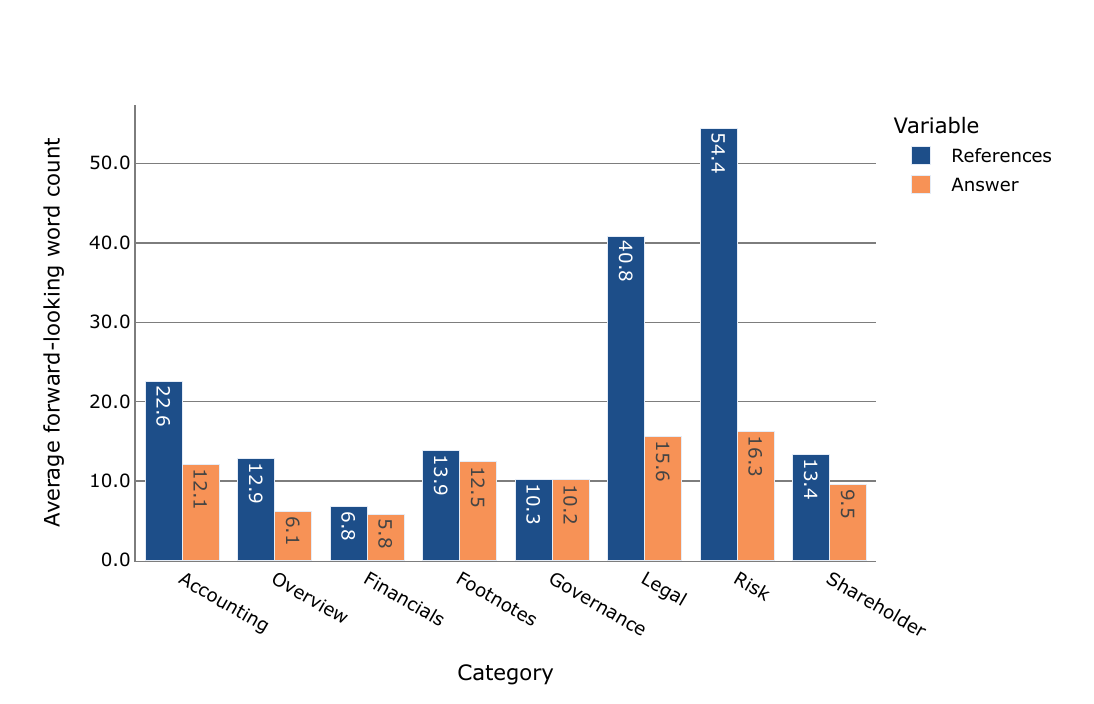}
    \caption{\textit{FinDER average forward-looking word count across answer and references by category.}}
\end{figure}

\subsection{Evaluation}

Evaluation of the underlying retrieval pipeline and the generation component is entirely delegated to LLMs by means of three specific metrics contained in the RAGAS open-source evaluation suite.\cite{es2025ragas}\footnote{Robustness to the evaluator model and to metrics specifications via prompt engineering are suggested as areas for further development in Section 5.}

Adherence of retrieved context to the original question is measured through \textit{Context Relevance}, which relies on two independent LLM judges that attribute each a score of either 0, 1 or 2, to be subsequently normalized and jointly averaged to return a number that ranges between 0 and 1 in steps of 0.25. An analogous methodology enables to compute \textit{Response Groundedness}, a metric that determines how well the generated output is adhering to the context retrieved in the initial step. The aforementioned metrics contribute to validating the steps that are necessary to extract the correct context from LSEG's database and generate the final answer based on that very context, akin to what would occur if context would stem from a standard RAG pipeline based entirely on document parsing and chunking rather than on MCP-wrapped tools, to conclude how extensively the generator model has utilized the provided context across answers, thus enabling discrimination between parametric and contextual knowledge.\cite{kandpal2023longtail,mallen2023parametrics,anthropic2024mcp,es2025ragas}

Finally, alignment between generated answers and ground truth is evaluated allowing for different degrees of tolerance to fairly reflect runtime-specific idiosyncrasies in the generated output that can result in differences, either format-wise or content-wise, when compared across runtimes to the available references. This is mostly motivated by the specific prompts that are used to determine \textit{Answer Accuracy} within RAGAS, and by the format of the ground truth stored in the dataset that is explored in the paper, which proves to be an issue mainly when assessing the quality of answers that contain purely quantitative figures, rather than more extensive textual explanations, normally handled better by LLM judges. A value of 1 is attributed to the generated answer if its content aligns satisfactorily to the ground truth, otherwise a value of 0 is assigned, determining a binary mapping that is referred to as \textit{Answer Accuracy}, similarly to the metric defined within RAGAS.\cite{es2025ragas}

Jointly considering these metrics enables not only to gauge how well the system has been able to perform on a given question, but also how grounded the generated answer is in the provided data, rather than in training knowledge that could incidentally yield the correct response, effectively working around the issue of \textit{long-tail knowledge}. In fact, it is worth mentioning that it becomes increasingly difficult to isolate the influence of knowledge injected into models during training by the same datasets that should serve as benchmarks. While an obvious solution to this issue is limiting the test set to those questions that lie beyond the training cutoff characterizing the model at hand, the available benchmarks would be heavily penalized in terms of size and variety. Instead, a proxy for gauging the severity of this problem without reducing the available datasets is, in the context of the MCP approach used here, testing the performance of the system with and without access to tools. If the model had been exposed to the specific answer to a question appearing in any dataset used for benchmarking, the model's tendency to refuse to answer or its hints to check the respective knowledge base allow to safely conclude that such exposure was not sufficient for the model to be able to generate a factually accurate answer as a result of token sampling. Therefore, the assumption that if the answer contains the data points extracted through tool calls then such an answer is comparable to the benchmark reference can stand, and this can be systematically assessed by means of \textit{Context Relevance}.\cite{kandpal2023longtail,anthropic2024mcp,huang2024hal,es2025ragas}

\section{Results}

The baseline model used for generation and evaluation is GPT-4o mini, and the parameters set for the purpose of this study are collected in Table 2 for reference. It is worth noting that the only parameter differing between the generator and evaluator instance is \textit{max\_completion\_tokens}, which is increased for the evaluator model to avoid hitting token limits while processing the underlying evaluation prompts.\cite{brown2020gpt3}

\begin{table}[h!]
\begin{center}
\caption{Baseline model parameters.}
\begin{tabular}{ccc}
    \hline
    \multirow{2}{*}{\textbf{Parameter}} &
    \multicolumn{2}{c}{\textbf{Value}} \\
    & \textbf{Generator} & \textbf{Evaluator} \\ \hline
    \textit{max\_completion\_tokens} & 2048 & 4096 \\
    \textit{seed} & 17 & 17 \\
    \textit{temperature} & 0.01 & 0.01 \\
    \textit{top\_p} & 0.15 & 0.15
\end{tabular}
\end{center}
\end{table}

Figure 4 relates the performance of the LSEG-powered LLM in addressing questions to the respective query category, as extracted from the preprocessed version of the reference dataset. It is immediately noticeable that questions concerning quantitative data are handled systematically better by the MCP-based architecture, with an unconditional average \textit{Answer Accuracy} of 69.7\% within the \textit{Financials} subset, as opposed to 50\% when jointly considering other question categories.\cite{anthropic2024mcp,choi2025finder,es2025ragas}\footnote{Benchmarking runs on smaller subsets of questions were performed restricting the baseline LLM from accessing any tool, confirming robustness of the LLM in computing answers from the provided context, rather than extracting them directly from memory. In fact, for the tested questions, the system would refuse to answer rather than hallucinating or providing an answer from memory, even and most importantly for data prior to the respective training cutoff.}

\begin{figure}[h!]
    \centering
    \includegraphics[width=.80\columnwidth]{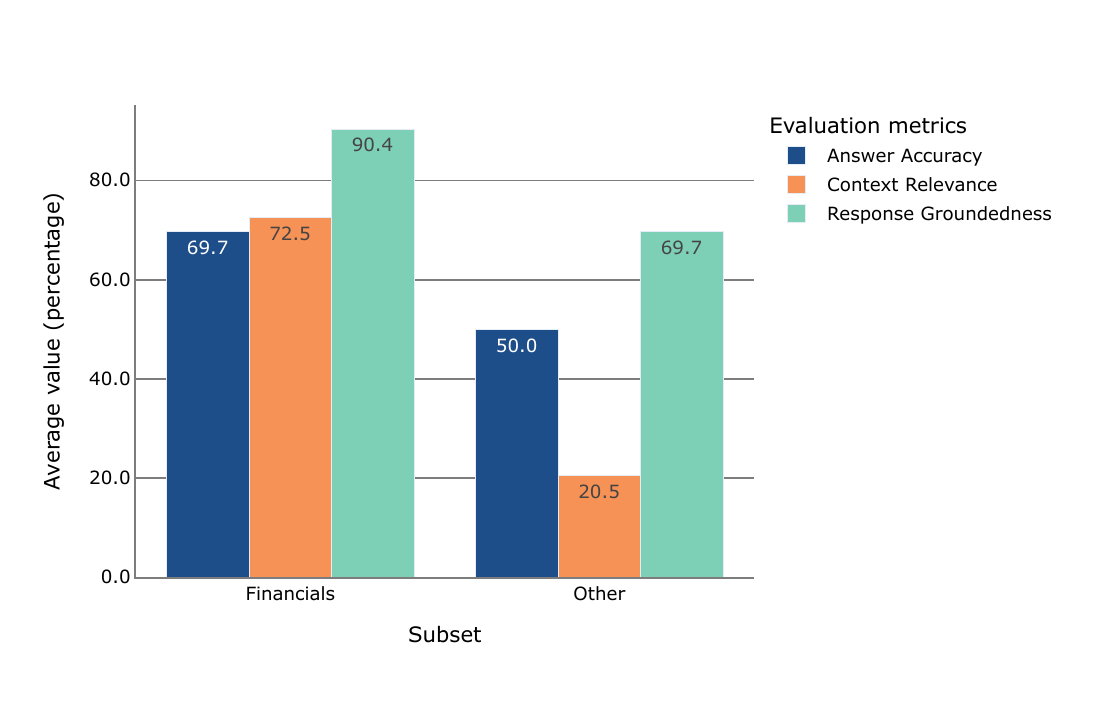}
    \caption{\textit{LSEG-powered LLM performance on FinDER by question category. Relative values computed over 990 and 4713 entries respectively for Financials and Other samples.}}
\end{figure}

To fully grasp the behavior of the constructed pipeline, Figure 4 also provides an overview of the \textit{Response Groundedness} scores across the reference dataset. The tendency is clear and once again characterized by higher values for questions concerning the extraction or processing of quantitative data, rather than the interpretation of quantitative data or inference based on qualitative data, with averages of 90.4\% and 69.7\%, respectively. Mechanically, high \textit{Response Groundedness} values are motivated by the extent to which LLMs are able to utilize the provided context, if any, to generate an answer. Conversely, it follows that an answer generated without context, or alternatively an answer that is generated by disregarding the provided context will lead to low \textit{Response Groundedness} values. Low \textit{Response Groundedness} scores have to be expected especially for those questions pertaining to qualitative data, considering that tools designed and developed for this study expose data points that are suitable to handle questions of purely quantitative nature, while questions of more qualitative nature frequently either lead to erroneous tool executions, or to the model answering without executing tools altogether.\cite{choi2025finder,es2025ragas}

Results are even more striking when conditioning by context quality, measured through \textit{Context Relevance}, as shown in Figure 5. Answers rooted in reliable evidence display an \textit{Answer Accuracy} of 73.8\% for the \textit{Financials} subset, whereas the valid counterpart extracted from other question categories only exhibits an alignment of 47.6\% with the reported ground truth. This evidence also draws attention to the developed infrastructure being able to recover highly relevant information for the majority of the quantitative questions characterizing the \textit{Financials} subset, as demonstrated by the conditioning step reducing the sample size by 32.6\% to 668 entries, as opposed to the drop of 86.5\% to 636 entries registered across the remaining question categories.\cite{choi2025finder,es2025ragas}

\begin{figure}[h!]
    \centering
    \includegraphics[width=.80\columnwidth]{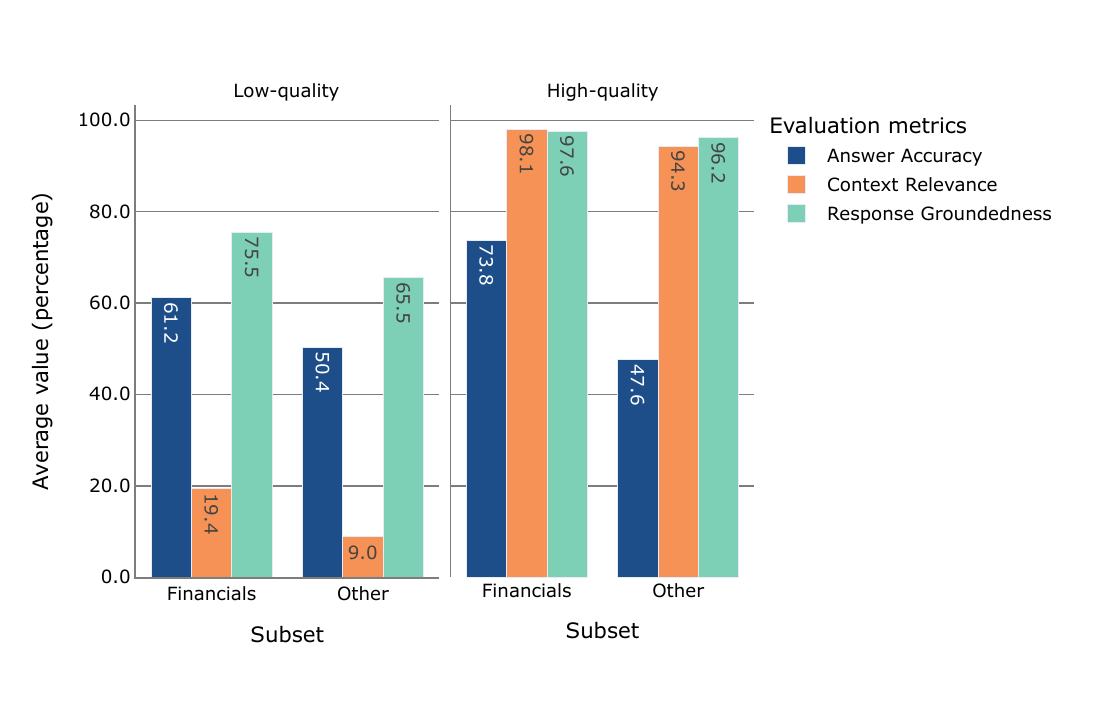}
    \caption{\textit{LSEG-powered LLM performance on FinDER by context quality and question category. An answer is defined as of high quality if the associated Context Relevance is equal or larger than 0.75. Relative values computed over 668 and 636 entries respectively for Financials and Other samples in the High-quality group, and over 322 and 4077 entries respectively for Financials and Other samples in the Low-quality group.}}
\end{figure}

Leveraging the more granular classification available that maps each question to the respective reasoning type allows to better assess how the system performs across task subtypes. The quality of context, as displayed in Figure 6, effectively drives system performance across all task types, with an average \textit{Answer Accuracy} of 75.3\% computed over 433 of the 577 answers belonging to the \textit{Numerical reasoning} group, as outlined in step 4 of the preprocessing procedure described in Section 3. The global maximum \textit{Answer Accuracy} stands at 80.4\% within the \textit{Compositional} task group, namely 204 of the 277 questions that are classified as involving multiple quantitative steps of potentially differing nature and representing the largest task group within the \textit{Numerical reasoning} set.\cite{choi2025finder,es2025ragas}

\begin{figure}[h!]
    \centering
    \includegraphics[width=.80\columnwidth]{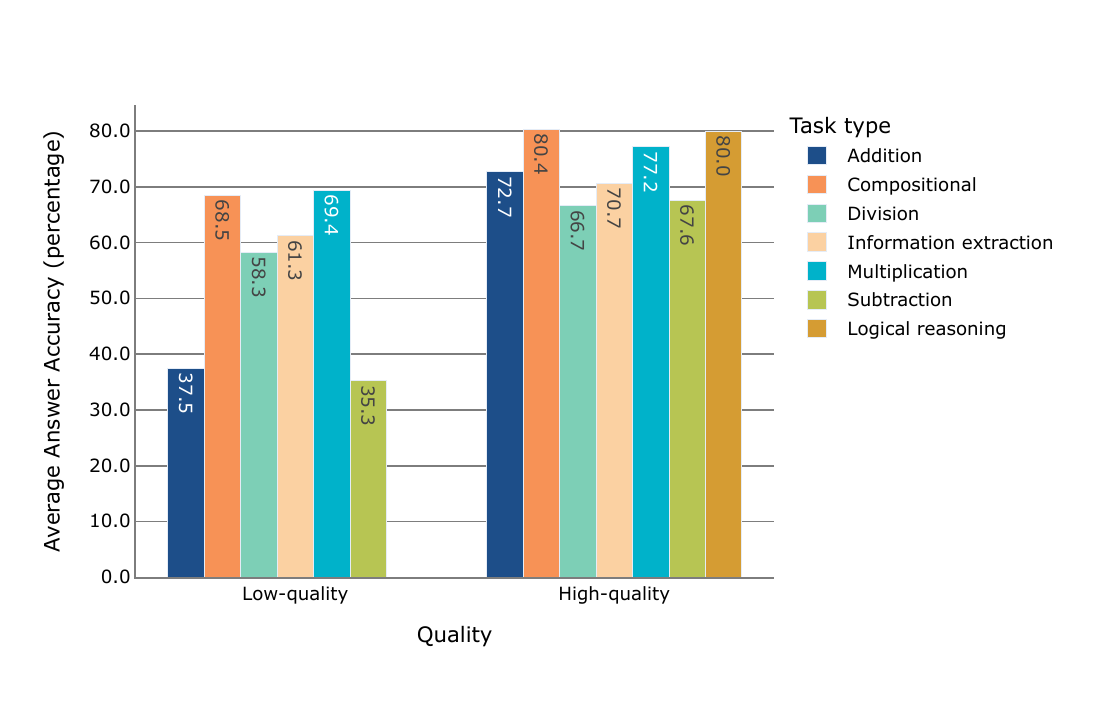}
    \caption{\textit{LSEG-powered LLM performance on FinDER Financials subset by context quality and task type. An answer is defined as of high quality if the associated Context Relevance is equal or larger than 0.75. Limited sample size reduces significance of results for Logical reasoning group.}}
\end{figure}

It is interesting to observe that a direct relationship exists between \textit{Answer Accuracy} and especially \textit{Context Relevance}, and the number occurrences characterizing reference answers, as visible in Figure 7. While, within the \textit{Financials} subset, the group of answers displaying no numbers configures a set of its own, both in terms of size and characterizing features, the trend is clear for the remaining entries: more numbers appearing in the reference answer imply a higher \textit{Context Relevance} and, in turn, a higher \textit{Answer Accuracy}. However, the impact of quantitative figures on informing the generated answer plateaus after 100 occurrences, potentially signaling that the generator model may be more prone to failing in filtering and capturing the most relevant information from then on, therefore injecting noise in the final output.\cite{choi2025finder,es2025ragas}\footnote{This may instead boil down to the choice of the generator model, with specific classes of models being more adept than others in selecting the right context to answer the question at hand.}

\begin{figure}[h!]
    \centering
    \includegraphics[width=.80\columnwidth]{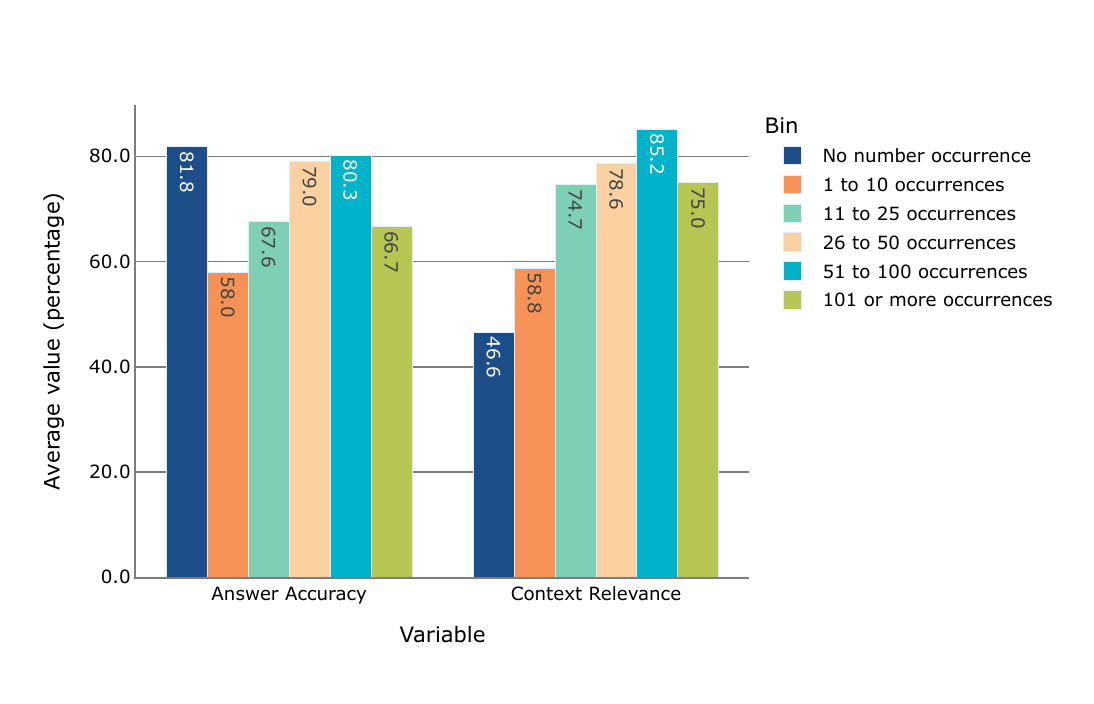}
    \caption{\textit{LSEG-powered LLM performance on FinDER Financials subset by number occurrence group and question type. Limited sample size reduces significance of results for "No number occurrences" and "101 or more occurrences" groups.}}
\end{figure}

\section{Conclusions}

The paper investigates the performance of a baseline LLM powered with access to data as provided by LSEG in order to answer questions collected within the FinDER dataset, following an MCP approach to orchestrate tool selection and execution. Further, the study provides substantial evidence that questions to be answered by means of quantitative data are of particular interest from the perspective of external data provider-powered agents, as they can often be answered without incurring in the overhead costs of establishing a standard RAG pipeline that renders documents containing the data necessary to answer readable by LLMs. This is particularly relevant to research, given that quantitative data is mostly collected in tabular form and standard RAG architectures require a sensible amount of work in order to parse tables correctly, while in the financial domain, data are already efficiently collected and disseminated by reliable sources. Performance of the system in terms of unconditional \textit{Answer Accuracy} averages at 69.7\% within the \textit{Financials} subset, whereas the score increases up to 80.4\% when uniquely considering \textit{High-quality} answers involving differing numerical calculations, proxied by a \textit{Context Relevance} equal or larger than 0.75, highlighting the effectiveness of MCP-based architectures in extracting and handling quantitative data for the purpose of financial QA.\cite{anthropic2024mcp,choi2025finder,es2025ragas}

A crucial aspect that unveils the lack of suitable datasets for further work in this direction, and thus highlights the necessity for human-curated benchmarks specifically designed for MCP-based solutions, is the performance of the approach outlined in the paper hinging on retrieving the same data points that are used in the respective benchmark datasets available for research. Given that the bulk of the effort in recent years has been revolving around document parsing and chunking, the typical question is shaped and tailored according to the nature of the knowledge base collected by the authors. For example, while a question that prompts the selection of a document chunk containing fiscal period headers from a knowledge base that only spans periods reported in a single financial report may well omit an indication of time, as the model will still be able to recover it from the chunk itself prior to generating the final answer, a good specification along all the defining variable dimensions of the question becomes fundamental when the knowledge base is defined dynamically, as is the case for the approach followed here.\cite{anthropic2024mcp}

Specific areas for development and limitations were identified while carrying out the study. With respect to evaluation, a relevant constraint lies in the choice and tweaking of evaluation prompts. These drive the system performance, as they can sensibly influence the resulting quantitative metrics. A certain degree of prompt tweaking is required to work around the idiosyncrasies of the dataset at hand and to prevent the evaluator model from labeling an answer as inaccurate due to minor formatting or rounding differences, and a delicate trade-off exists when attempting to minimize false negatives and false positives in the context of a LLM-as-a-Judge pipeline. More research in this area is therefore necessary to conclude what methods work best for a specific type of dataset, along with a systematic assessment of the influence of tool execution and answer generation on latency and token consumption, two parameters that are critical when aiming at scaling and turning the pipeline into a production-ready solution. This would serve to better quantify the added value of following an MCP approach rather than relying on standard RAG architectures.\cite{anthropic2024mcp,es2025ragas}

Building on the topic of evaluation, further opportunities for development also lie in defining custom metrics that on the one hand better leverage the potential of available LLMs, and on the other hand more precisely capture each and every dimension of the pipeline where bias may be injected by construction or due to inference. Strictly related to the limitation posed to the study by the evaluation step, an area for development can be found in defining a coherent set of rules to be followed when defining the evaluation prompts necessary for assessing generated answers.\footnote{As an example, dynamic tolerance bands may penalize an answer as the absolute value of its key variable increases, such that a rounding error of 0.1 may be ignored if the variable is a percentage beyond $\pm0.5$, or of monetary type and expressed in units of currency, but at the same time lead to the answer being classified as inaccurate if the variable is expressed in billions, or a percentage within $\pm0.5$.}

Assessing robustness of the system to the choice of the generator and evaluator models also represents an important area for future research, given that fully relying on a single model provider may inject bias, most importantly in the evaluation phase. At the same time, a robustness analysis on the impact over system performance of model parameters such as temperature or maximum output tokens would strengthen the conclusions reached in this study. Finally, with respect to robustness, testing how the system performs as data providers within and outside the commercial landscape change would considerably enrich the results determined in the study.

Finally, the design of further tools and of an architecture that enables their sequential or parallel execution to effectively and efficiently retrieve data points relevant to answer user questions constitutes a key area for development, considering that the high quality of data available in the financial industry shifts focus on the issue of optimizing retrieval through the most suitable API endpoints, rather than on data availability and validation. Moving in this direction, creating a set of tools that reliably compute more complex financial ratios and that are chained to tools retrieving the necessary data points represents an initial way to reduce model hallucination, thereby improving the aforementioned evaluation metrics.

\clearpage

\bibliographystyle{unsrt}
\bibliography{paperbibliography}

@article{altman1968bankrupt,
    title={{Financial Ratios, Discriminant Analysis and the Prediction of Corporate Bankruptcy.}},
    author={Edward I. Altman},
    journal={The Journal of Finance},
    volume ={23},
    number ={4},
    pages ={589--609},
    year={1968},
    DOI={https://doi.org/10.2307/2978933}
}

@article{altman1984business,
    title={{The success of business failure prediction models: An international survey.}},
    author={Edward I. Altman},
    journal={Journal of Banking and Finance},
    volume ={8},
    number ={2},
    pages ={171--198},
    year={1984},
    DOI={https://doi.org/10.1016/0378-4266(84)90003-7}
}

@article{almamy2016zscore,
    title={{An evaluation of Altman's Z-score using cash flow ratio to predict corporate failure amid the recent financial crisis: Evidence from the UK.}},
    author={Jeehan Almamy{,} John Aston{,} Leonard N. Ngwa},
    journal={Journal of Corporate Finance},
    volume ={36},
    number ={4},
    pages ={278--285},
    year={2016},
    DOI={https://doi.org/10.1016/j.jcorpfin.2015.12.009}
}

@article{bahdanau2014attention,
    title={{Neural Machine Translation by Jointly Learning to Align and Translate.}},
    author={Dzmitry Bahdanau{,} Kyunghyun Cho{,} Yoshua Bengio},
    journal={arXiv preprint arXiv:1409.0473},
    year={2014},
    DOI={https://doi.org/10.48550/arXiv.1409.0473}
}

@article{barth2022nonanswers,
    title={{“Let me get back to you” - A machine learning approach to measuring non-answers.}},
    author={Andreas Barth{,} Sasan Mansouri{,} Fabian Wöbbeking},
    journal={Management Science},
    volume ={69},
    number ={10},
    pages ={6333--6348},
    year={2022},
    DOI={https://doi.org/10.1287/mnsc.2022.4597}
}

@article{bhandari2025mcp,
    title={{Multi-Scale Network Dynamics and Systemic Risk: A Model Context Protocol Approach to Financial Markets.}},
    author={Avishek Bhandari},
    journal={arXiv preprint arXiv:2507.08065},
    year={2025},
    DOI={https://doi.org/10.48550/arXiv.2507.08065}
}

@article{beaver2012accounting,
    title={{Do differences in financial reporting attributes impair the predictive ability of financial ratios for bankruptcy?}},
    author={William H. Beaver{,} Maria Correia{,} Maureen F. McNichols},
    journal={Review of Accounting Studies},
    volume ={17},
    number ={4},
    pages ={969--1010},
    year={2012},
    DOI={https://doi.org/10.1007/s11142-012-9186-7}
}

@article{bozanic2018forward,
    title={{Management earnings forecasts and other forward-looking statements.}},
    author={Zahn Bozanic{,} Darren T. Roulstone{,} Andrew Van Buskirk},
    journal={Journal of Accounting and Economics},
    volume ={65},
    number ={1},
    pages ={1--20},
    year={2018},
    DOI={https://doi.org/10.1016/j.jacceco.2017.11.008}
}

@article{brown2020gpt3,
    title={{Language Models are Few-Shot Learners.}},
    author={Tom B. Brown{,} Benjamin Mann{,} Nick Ryder{,} Melanie Subbiah{,} Jared Kaplan{,} Prafulla Dhariwal{,} Arvind Neelakantan{,} Pranav Shyam{,} Girish Sastry{,} Amanda Askell{,} Sandhini Agarwal{,} Ariel Herbert-Voss{,} Gretchen Krueger{,} Tom Henighan{,} Rewon Child{,} Aditya Ramesh{,} Daniel M. Ziegler{,} Jeffrey Wu{,} Clemens Winter{,} Christopher Hesse{,} Mark Chen{,} Eric Sigler{,} Mateusz Litwin{,} Scott Gray{,} Benjamin Chess{,} Jack Clark{,} Christopher Berner{,} Sam McCandlish{,} Alec Radford{,} Ilya Sutskever{,} Dario Amodei},
    journal={arXiv preprint arXiv:2005.14165},
    year={2020},
    DOI={https://doi.org/10.48550/arXiv.2005.14165}
}

@article{chen2021finqa,
    title={{FinQA: A Dataset of Numerical Reasoning over Financial Data.}},
    author={Zhiyu Chen{,} Wenhu Chen{,} Charese Smiley{,} Sameena Shah{,} Iana Borova{,} Dylan Langdon{,} Reema Moussa{,} Matt Beane{,} Ting-Hao Huang{,} Bryan Routledge{,} William Yang Wang},
    journal={arXiv preprint arXiv:2109.00122},
    year={2021},
    DOI={https://doi.org/10.48550/arXiv.2109.00122}
}

@article{choi2025finder,
    title={{FinDER: Financial Dataset for Question Answering and Evaluating Retrieval-Augmented Generation.}},
    author={Chanyeol Choi{,} Jihoon Kwon{,} Jaeseon Ha{,} Hojun Choi{,} Chaewoon Kim{,} Yongjae Lee{,} Jy-yong Sohn{,} Alejandro Lopez-Lira},
    journal={arXiv preprint arXiv:2504.15800},
    year={2025},
    DOI={https://doi.org/10.48550/arXiv.2504.15800}
}

@article{corradaemmanuel2025llmasjudge,
    title={{No-Knowledge Alarms for Misaligned LLMs-as-Judges.}},
    author={Andrés Corrada-Emmanuel},
    journal={arXiv preprint arXiv:2509.08593},
    year={2025},
    DOI={https://doi.org/10.48550/arXiv.2509.08593}
}

@article{deangelo1994accounting,
    title={{Accounting choice in troubled companies.}},
    author={Harry DeAngelo{,} Linda DeAngelo{,} Douglas J. Skinner},
    journal={Journal of Accounting and Economics},
    volume ={17},
    number ={1},
    pages ={113--143},
    year={1994},
    DOI={https://doi.org/10.1016/0165-4101(94)90007-8}
}

@article{dechow2012earnings,
    title={{Detecting Earnings Management: A New Approach.}},
    author={Patricia M. Dechow{,} Amy P. Hutton{,} Jung H. Kim{,} Richard G. Sloan},
    journal={Journal of Accounting Research},
    volume ={50},
    number ={2},
    pages ={275--334},
    year={2012},
    DOI={https://doi.org/10.1111%2Fj.1475-679X.2012.00449.x}
}

@article{errico2025mcp,
    title={{Securing the Model Context Protocol (MCP): Risks, Controls, and Governance.}},
    author={Herman Errico{,} Jiquan Ngiam{,} Shanita Sojan},
    journal={arXiv preprint arXiv:2511.20920},
    year={2025},
    DOI={https://doi.org/10.48550/arXiv.2511.20920}
}

@article{es2025ragas,
    title={{Ragas: Automated Evaluation of Retrieval Augmented Generation.}},
    author={Shahul Es{,} Jithin James{,} Luis Espinosa-Anke{,} Steven Schockaert},
    journal={arXiv preprint arXiv:2309.15217},
    year={2025},
    DOI={https://doi.org/10.48550/arXiv.2309.15217}
}

@article{ferrazzi2026rag,
    title={{Is Agentic RAG worth it? An experimental comparison of RAG approaches.}},
    author={Pietro Ferrazzi{,} Milica Cvjeticanin{,} Alessio Piraccini{,} Davide Giannuzzi},
    journal={arXiv preprint arXiv:2601.07711},
    year={2026},
    DOI={https://doi.org/10.48550/arXiv.2601.07711}
}

@article{franz2014earnings,
    title={{Impact of proximity to debt covenant violation on earnings management.}},
    author={Diana R. Franz{,} Hassan R. HassabElnaby{,} Gerald J. Lobo},
    journal={Review of Accounting Studies},
    volume ={19},
    number ={},
    pages ={473--505},
    year={2014},
    DOI={https://doi.org/10.1007/s11142-013-9252-9}
}

@article{gao2024rag,
    title={{Retrieval-Augmented Generation for Large Language Models: A Survey.}},
    author={Yunfan Gao{,} Yun Xiong{,} Xinyu Gao{,} Kangxiang Jia{,} Jinliu Pan{,} Yuxi Bi{,} Yi Dai{,} Jiawei Sun{,} Meng Wang{,} Haofen Wang},
    journal={arXiv preprint arXiv:2312.10997},
    year={2024},
    DOI={https://doi.org/10.48550/arXiv.2312.10997}
}

@article{graham2005reporting,
    title={{The economic implications of corporate financial reporting.}},
    author={John R. Graham{,} Campbell R. Harvey{,} Shiva Rajgopal},
    journal={Journal of Accounting and Economics},
    volume ={40},
    number ={1--3},
    pages ={3--73},
    year={2005},
    DOI={https://doi.org/10.1016/j.jacceco.2005.01.002}
}

@article{guo2025mcp,
    title={{MCP-AgentBench: Evaluating Real-World Language Agent Performance with MCP-Mediated Tools.}},
    author={Zikang Guo{,} Benfeng Xu{,} Chiwei Zhu{,} Wentao Hong{,} Xiaorui Wang{,} Zhendong Mao},
    journal={arXiv preprint arXiv:2509.09734},
    year={2025},
    DOI={https://doi.org/10.48550/arXiv.2509.09734}
}

@article{gu2025llmasjudge,
    title={{A Survey on LLM-as-a-Judge.}},
    author={Jiawei Gu{,} Xuhui Jiang{,} Zhichao Shi{,} Hexiang Tan{,} Xuehao Zhai{,} Chengjin Xu{,} Wei Li{,} Yinghan Shen{,} Shengjie Ma{,} Honghao Liu{,} Saizhuo Wang{,} Kun Zhang{,} Yuanzhuo Wang{,} Wen Gao{,} Lionel Ni{,} Jian Guo},
    journal={arXiv preprint arXiv:2411.15594},
    year={2025},
    DOI={https://doi.org/10.48550/arXiv.2411.15594}
}

@article{healy1985bonus,
    title={{The effect of bonus schemes on accounting decisions.}},
    author={Paul Healy},
    journal={Journal of Accounting and Economics},
    volume ={7},
    number ={1--3},
    pages ={85--107},
    year={1985},
    DOI={https://doi.org/10.1016/0165-4101(85)90029-1}
}

@article{hou2025mcp,
    title={{Model Context Protocol (MCP): Landscape, Security Threats, and Future Research Directions.}},
    author={Xinyi Hou{,} Yanjie Zhao{,} Shenao Wang{,} Haoyu Wang},
    journal={arXiv preprint arXiv:2503.23278},
    year={2025},
    DOI={https://doi.org/10.48550/arXiv.2503.23278}
}

@article{huang2024hal,
    title={{A Survey on Hallucination in Large Language Models: Principles, Taxonomy, Challenges, and Open Questions.}},
    author={Lei Huang{,} Weijiang Yu{,} Weitao Ma{,} Weihong Zhong{,} Zhangyin Feng{,} Haotian Wang{,} Qianglong Chen{,} Weihua Peng{,} Xiaocheng Feng{,} Bing Qin{,} Ting Liu},
    journal={arXiv preprint arXiv:2311.05232},
    year={2024},
    DOI={https://doi.org/10.48550/arXiv.2311.05232}
}

@article{islam2023finbench,
    title={{FinanceBench: A New Benchmark for Financial Question Answering.}},
    author={Pranab Islam{,} Anand Kannappan{,} Douwe Kiela{,} Rebecca Qian{,} Nino Scherrer{,} Bertie Vidgen},
    journal={arXiv preprint arXiv:2311.11944},
    year={2023},
    DOI={https://doi.org/10.48550/arXiv.2311.11944}
}

@article{kandpal2023longtail,
    title={{Large Language Models Struggle to Learn Long-Tail Knowledge.}},
    author={Nikhil Kandpal{,} Haikang Deng{,} Adam Roberts{,} Eric Wallace{,} Colin Raffel},
    journal={arXiv preprint arXiv:2211.08411},
    year={2023},
    DOI={https://doi.org/10.48550/arXiv.2211.08411}
}

@article{kim2017attention,
    title={{Structured Attention Networks.}},
    author={Yoon Kim{,} Carl Denton{,} Luong Hoang{,} Alexander M. Rush},
    journal={arXiv preprint arXiv:1702.00887},
    year={2017},
    DOI={https://doi.org/10.48550/arXiv.1702.00887}
}

@article{levy2024bias,
    title={{Caution Ahead: Numerical Reasoning and Look-ahead Bias in AI Models.}},
    author={Bradford L. Levy},
    journal={Available at SSRN: https://ssrn.com/abstract=5082861},
    year={2024},
    DOI={http://dx.doi.org/10.2139/ssrn.5082861}
}

@article{lewis2020rag,
    title={{Retrieval-Augmented Generation for Knowledge-Intensive NLP Tasks.}},
    author={Patrick Lewis{,} Ethan Perez{,} Aleksandara Piktus{,} Fabio Petroni{,} Vladimir Karpukhin{,} Naman Goyal{,} Heinrich Küttler{,} Mike Lewis{,} Wen-tau Yih{,} Tim Rocktäschel{,} Sebastian Riedel{,} Douwe Kiela},
    journal={arXiv preprint arXiv:2005.11401},
    year={2020},
    DOI={https://doi.org/10.48550/arXiv.2005.11401}
}

@article{li2025llmasjudge,
    title={{Who Judges the Judge? LLM Jury-on-Demand: Building Trustworthy LLM Evaluation Systems.}},
    author={Xiaochuan Li{,} Ke Wang{,} Girija Gouda{,} Shubham Choudhary{,} Yaqun Wang{,} Linwei Hu{,} Joel Vaughan{,} Freddy Lecue},
    journal={arXiv preprint arXiv:2512.01786},
    year={2025},
    DOI={https://doi.org/10.48550/arXiv.2512.01786}
}

@article{mallen2023parametrics,
    title={{When Not to Trust Language Models: Investigating Effectiveness of Parametric and Non-Parametric Memories.}},
    author={Alex Mallen{,} Akari Asai{,} Victor Zhong{,} Rajarshi Das{,} Daniel Khashabi{,} Hannaneh Hajishirzi},
    journal={arXiv preprint arxiv:2212.10511},
    year={2023},
    DOI={https://doi.org/10.48550/arXiv.2212.10511}
}

@article{sarkar2024bias,
    title={{Lookahead Bias in Pretrained Language Models.}},
    author={Suproteem K. Sarkar{,} Keyon Vafa},
    journal={Available at SSRN: https://ssrn.com/abstract=4754678},
    year={2024},
    DOI={http://dx.doi.org/10.2139/ssrn.4754678}
}

@article{sarmah2023rag,
    title={{Towards reducing hallucination in extracting information from financial reports using Large Language Models.}},
    author={Bhaskarjit Sarmah{,} Tianjie Zhu{,} Dhagash Mehta{,} Stefano Pasquali},
    journal={arXiv preprint arXiv:2310.10760},
    year={2023},
    DOI={https://doi.org/10.48550/arXiv.2310.10760}
}

@article{setty2024rag,
    title={{Improving Retrieval for RAG based Question Answering Models on Financial Documents.}},
    author={Spurthi Setty{,} Harsh Thakkar{,} Alyssa Lee{,} Eden Chung{,} Natan Vidra},
    journal={arXiv preprint arXiv:2404.07221},
    year={2024},
    DOI={https://doi.org/10.48550/arXiv.2404.07221}
}

@article{tan2025table,
    title={{Understanding Structured Financial Data with LLMs: A Case Study on Fraud Detection.}},
    author={Xuwei Tan{,} Yao Ma{,} Xueru Zhang},
    journal={arXiv preprint arxiv:2512.13040},
    year={2025},
    DOI={https://doi.org/10.48550/arXiv.2512.13040}
}

@article{vaswani2017att,
    title={{Attention Is All You Need.}},
    author={Ashish Vaswani{,} Noam Shazeer{,} Niki Parmar{,} Jakob Uszkoreit{,} Llion Jones{,} Aidan N. Gomez{,} Lukasz Kaiser{,} Illia Polosukhin},
    journal={arXiv preprint arXiv:1706.03762},
    year={2017},
    DOI={https://doi.org/10.48550/arXiv.1706.03762}
}

@article{wang2025mcpbench,
    title={{MCP-Bench: Benchmarking Tool-Using LLM Agents with Complex Real-World Tasks via MCP Servers.}},
    author={Zhenting Wang{,} Qi Chang{,} Hemani Patel{,} Shashank Biju{,} Cheng-En Wu{,} Quan Liu{,} Aolin Ding{,} Alireza Rezazadeh{,} Ankit Shah{,} Yujia Bao{,} Eugene Siow},
    journal={arXiv preprint arXiv:2508.20453},
    year={2025},
    DOI={https://doi.org/10.48550/arXiv.2508.20453}
}

@article{wang2025mcpguard,
    title={{MCPGuard : Automatically Detecting Vulnerabilities in MCP Servers.}},
    author={Bin Wang{,} Zexin Liu{,} Hao Yu{,} Ao Yang{,} Yenan Huang{,} Jing Guo{,} Huangsheng Cheng{,} Hui Li{,} Huiyu Wu},
    journal={arXiv preprint arXiv:2510.23673},
    year={2025},
    DOI={https://doi.org/10.48550/arXiv.2510.23673}
}

@article{yepes2024chunk,
    title={{Financial Report Chunking for Effective Retrieval Augmented Generation.}},
    author={Antonio Jimeno Yepes{,} Yao You{,} Jan Milczek{,} Sebastian Laverde{,} Renyu Li},
    journal={arXiv preprint arXiv:2402.05131},
    year={2024},
    DOI={https://doi.org/10.48550/arXiv.2402.05131}
}

@article{zeng2025mcp,
    title={{QuantMCP: Grounding Large Language Models in Verifiable Financial Reality.}},
    author={Yifan Zeng},
    journal={arXiv preprint arXiv:2506.06622},
    year={2025},
    DOI={https://doi.org/10.48550/arXiv.2506.06622}
}

@article{zhu2021tatqa,
    title={{TAT-QA: A Question Answering Benchmark on a Hybrid of Tabular and Textual Content in Finance.}},
    author={Fengbin Zhu{,} Wenqiang Lei{,} Youcheng Huang{,} Chao Wang{,} Shuo Zhang{,} Jiancheng Lv{,} Fuli Feng{,} Tat-Seng Chua},
    journal={arXiv preprint arXiv:2105.07624},
    year={2021},
    DOI={https://doi.org/10.48550/arXiv.2105.07624}
}

@article{zhuge2024llmasjudge,
    title={{Agent-as-a-Judge: Evaluate Agents with Agents.}},
    author={Mingchen Zhuge{,} Changsheng Zhao{,} Dylan Ashley{,} Wenyi Wang{,} Dmitrii Khizbullin{,} Yunyang Xiong{,} Zechun Liu{,} Ernie Chang{,} Raghuraman Krishnamoorthi{,} Yuandong Tian{,} Yangyang Shi{,} Vikas Chandra{,} Jürgen Schmidhuber},
    journal={arXiv preprint arXiv:2410.10934},
    year={2024},
    DOI={https://doi.org/10.48550/arXiv.2410.10934}
}

@misc{anthropic2024mcp,
    author={Anthropic Team},
    title={{Introducing the Model Context Protocol.}},
    howpublished="\url{https://www.anthropic.com/news/model-context-protocol}",
    year={2024},
}

@misc{hussain2025lseg,
    author={London Stock Exchange Group},
    title={{Scaling AI in Financial Services with LSEG’s Trusted AI Ready Content and Model Context Protocol (MCP).}},
    howpublished="\url{https://www.lseg.com/en/insights/scaling-ai-financial-services-with-lseg-trusted-ai-ready-content-mcp}",
    year={2025},
}

@misc{challapally2025nanda,
    author={Aditya Challapally{,} Chris Pease{,} Ramesh Raskar{,} Pradyumna Chari},
    title={{The GenAI Divide State of AI in Business 2025.}},
    howpublished="\url{https://mlq.ai/media/quarterly_decks/v0.1_State_of_AI_in_Business_2025_Report.pdf}",
    year={2025}}

\clearpage

\appendix
\section*{Appendix}
\renewcommand{\thesubsection}{\Alph{subsection}}

\subsection{Extended results}

\setcounter{figure}{0}
\renewcommand{\thefigure}{A.\arabic{figure}}

The relationship highlighted in Figure 7 is also visible when conditioning based on the resulting \textit{Context Relevance} to distinguish \textit{High-quality} answers from the overall sample. Figure A.1 depicts exactly this evidence and contributes to purging the effect of hallucination from other categories that would otherwise exhibit a misleadingly high \textit{Answer Accuracy}.

\begin{figure}[h!]
    \centering
    \includegraphics[width=.80\columnwidth]{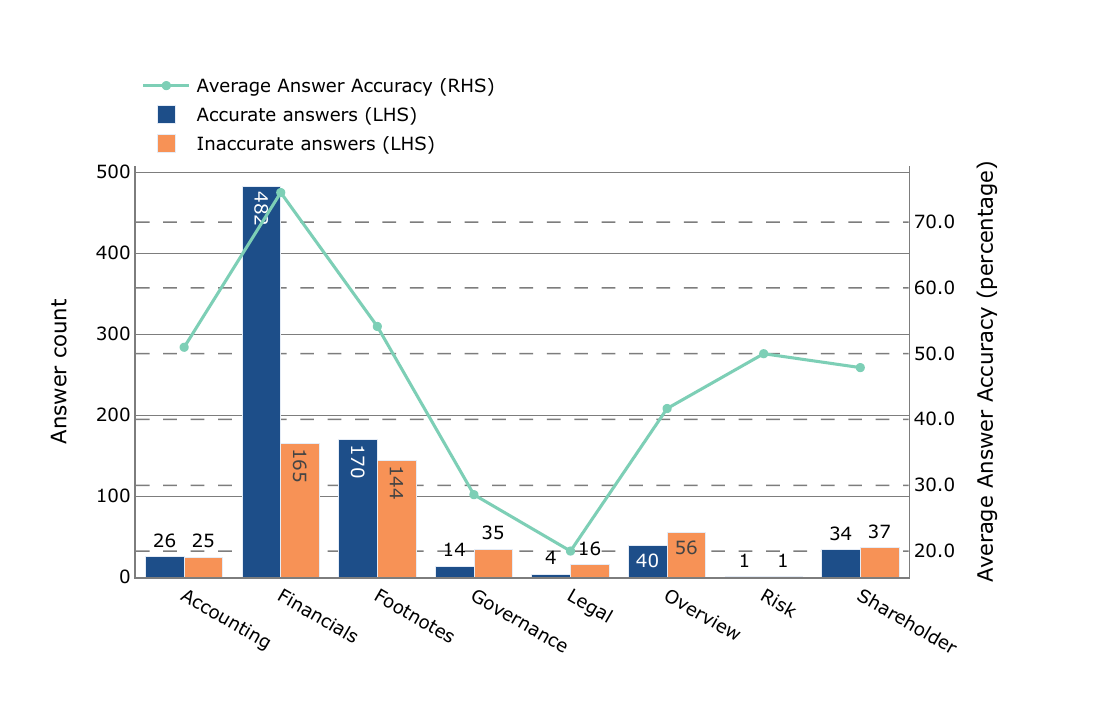}
    \caption{\textit{FinDER High-quality answer count (LHS) vs. average Answer Accuracy (RHS) by category.}}
\end{figure}

A more granular view is displayed for \textit{Context Relevance} and \textit{Response Groundedness} in Figure A.2, which shows that the \textit{Financials} subset is characterized by the largest number of answers displaying the highest quality, both in the retrieval and in the generation components of the pipeline, pushing in turn the aggregate averages depicted in Figure 4 up.

\begin{figure}[h!]
    \centering
    \includegraphics[width=.80\columnwidth]{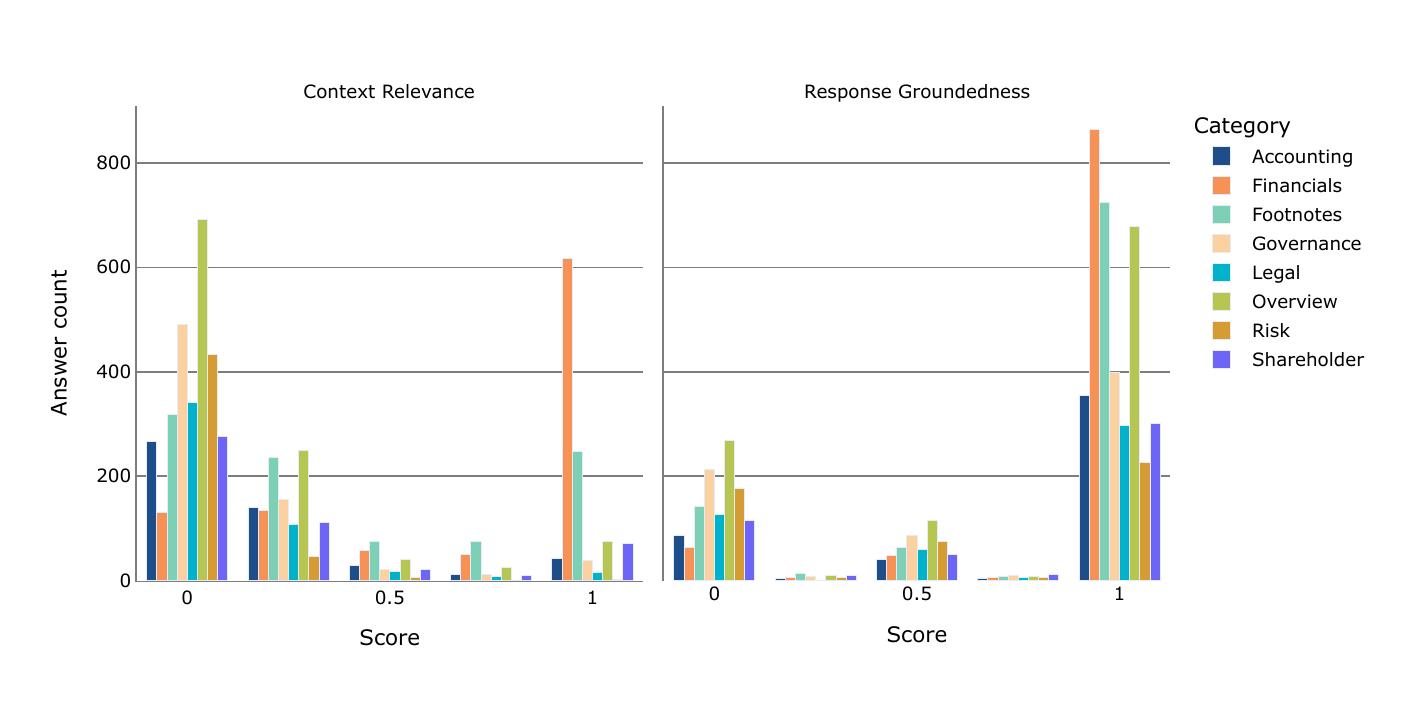}
    \caption{\textit{FinDER Context Relevance (LHS) vs. Response Groundedness (RHS) answer count by category.}}
\end{figure}

In the same fashion, Figure A.3 showcases the relationship already outlined in Figure 7 and puts it in perspective across FinDER categories, for which generally no clear pattern can be found, except for the \textit{Financials} subset.

\begin{figure}[h!]
    \centering
    \includegraphics[width=.80\columnwidth]{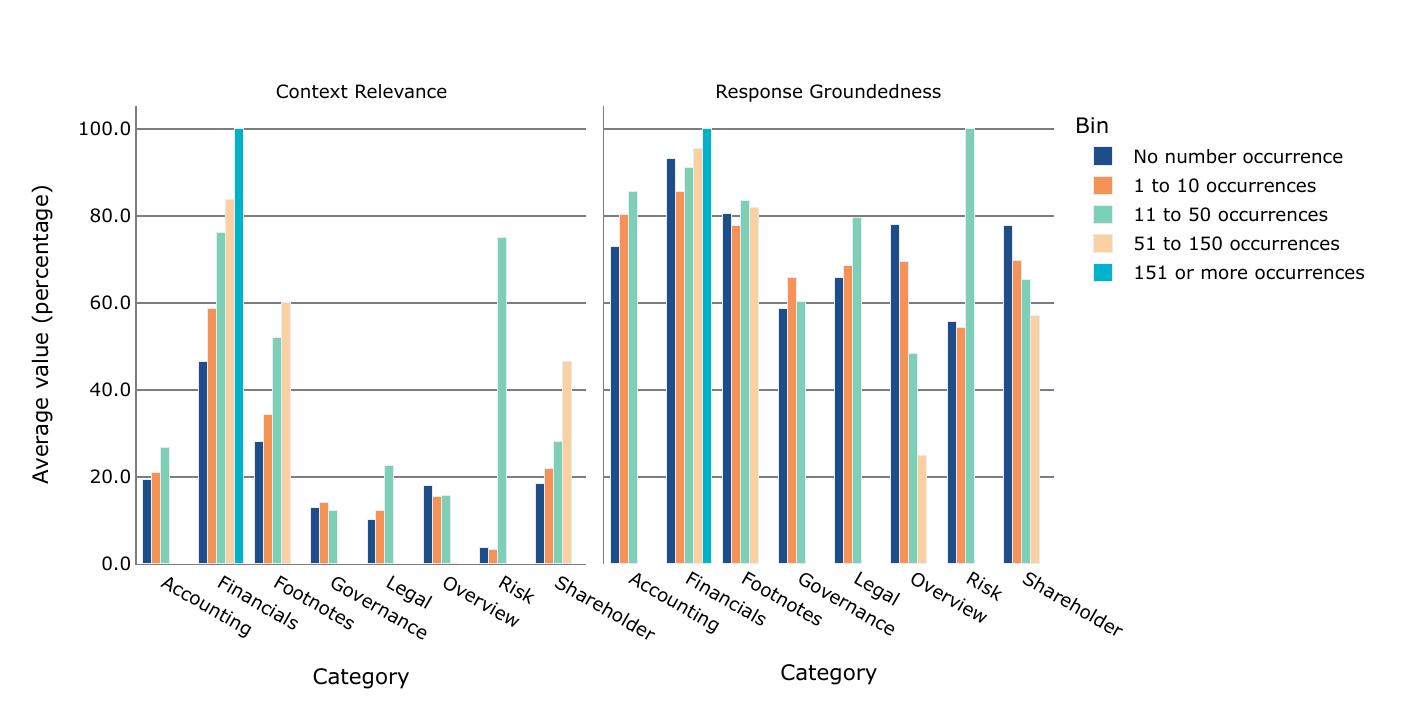}
    \caption{\textit{FinDER average Context Relevance (LHS) vs. average Response Groundedness (RHS) across number occurrences by category.}}
\end{figure}

\subsection{Backend prompts}

The prompts that are defined in the backend to enable benchmarking and evaluation are collected below for convenience:

\begin{itemize}
    \item \textit{Benchmarking:} "You are a helpful assistant. If you receive a question without any indication of time period, always assume it to refer to FY2023. Always limit tool executions to 5 fiscal periods prior to the reference one."
    \item \textit{Evaluation:} No system prompt explicitly defined.
    \item \textit{Answer Accuracy:} "The response is inaccurate if it is incorrect and fails to address any aspect of the reference. The response is accurate if it displays the same or partially the same data reported in the reference. Rounding and formatting differences are allowed."
    \item \textit{Context Relevance:} Default RAGAS prompts.
    \item \textit{Response Groundedness:} Default RAGAS prompts.
\end{itemize}

The rationale behind the phrasing of these prompts is that the majority of the financial reports forming the knowledge base for the reference dataset refer to fiscal year 2023, and limiting tool execution to five fiscal periods prior to the reference one lowers the likelihood of incurring into API request limits from LSEG, while at the same time preventing instances in which the model attempts to execute tools indefinitely. Given the flexibility required to cover all the questions contained within the reference dataset, it was chosen not to rely on model parameters such as \textit{max\_tool\_calls} that would otherwise represent the optimal option for this type of issue.

Allowing for full or partial coverage of the data points displayed in the respective ground truth captures the frequent occurrences where the system-generated answer is labeled as inaccurate due to not discussing all the data points present in the ground truth, even when all the data points are correctly extracted through tool execution. This is especially problematic given that the parameters used to generate the ground truth are unknown and can easily lead to generating answers that develop along different dimensions as opposed to those covered by the LSEG-powered system, whose parameters are reported in Section 4.

\end{document}